\documentclass[aps,showpacs,twocolumn,floatfix,amsmath,amssymb,nofootinbib]{revtex4-1}
\usepackage{amssymb}
\usepackage{amsmath}
\usepackage{epsfig}
\usepackage{graphicx}
\usepackage{latexsym}
\usepackage{bm,latexsym}
\usepackage{mathrsfs}
\usepackage{float}
\usepackage{pbox}

\def\Journal#1#2#3#4{{#1} {\bf #2}, #3 (#4)}

\def\AJ{{Astron. J.}}

\def\APJ{{Astrophys. J.}}

\def\GRG{{Gen. Relativ. Gravit.}}
\def\IJMPD{{Int. Jour. Mod. Phys. D}}
\def\JCAP{{JCAP}}

\def\LRR{{Living Rev. Rel.}} 
\def\MNRAS{{Mon. Not. Roy. Astron. Soc.}} 
\def\NAT{{Nature}}

\def\PLB{{Phys. Lett.}  B}

\def\PRL{Phys. Rev. Lett.}

\def\PRD{{Phys. Rev.} D}
\def\PR{{Phys. Rep.}}
\def\PTP{{Prog. Theor. Phys.}}

\def\RMP{{Rev. Mod. Phys.}}

\def\MNRAS{{Mon. Not. R. Astron. Soc.}}

\begin{document}


\title{Note on the equation of state of geometric dark-energy in $f(R)$ gravity}

\author{Luisa G. Jaime$^1$}
\email{luisa@nucleares.unam.mx}

\author{Leonardo Pati\~no$^2$}
\email{leopj@ciencias.unam.mx}

\author{Marcelo Salgado$^1$}
\email{marcelo@nucleares.unam.mx}

\affiliation{$^1$Instituto de Ciencias Nucleares, Universidad Nacional
Aut\'onoma de M\'exico, A.P. 70-543, M\'exico D.F. 04510, M\'exico \\
$^2$ Facultad de Ciencias, Universidad Nacional
Aut\'onoma de M\'exico, A.P. 50-542, M\'exico D.F. 04510, M\'exico }


\date{\today}

    
\begin{abstract}
We focus on the analysis of three inequivalent equations of state of geometric 
dark energy in $f(R)$ cosmology that have been considered in the past and discuss their differences, advantages and drawbacks.
\end{abstract}




\maketitle


\section{Introduction}

Astronomical observations based on type Ia supernovae (SNIa) together with the assumption that the Universe is homogeneous and isotropic at large scales led to 
the conclusion that the Universe is currently expanding in an accelerated way~\cite{Perlmutter1999,Riess1998,Amanullah2010}. This phenomenon can be explained 
by appealing to the existence of a cosmological constant  $\Lambda$. However, despite the simplicity and success of this paradigm, several theoretical as well 
as epistemological arguments have been put forward as objections against such a simple model of the Universe (see Refs.~\cite{Bianchi2010} for a discussion).

In order to provide an alternative explanation for the acceleration of the Universe, and even for dark matter, several modified theories of gravity (MTG) have 
been proposed. Perhaps the most popular MTG over the past ten years and the one we focus in this article are $f(R)$ metric 
theories~\cite{f(R),Capozziello2008a,Sotiriou2010,deFelice2010}, where an {\it a priori} arbitrary function of the Ricci scalar $R$ replaces $R$ itself in 
the gravitational Lagrangian. 

In the present article we analyze the associated equation of state (EOS) of the geometric dark energy (GDE) in $f(R)$ gravity and we show that, among the 
several definitions used in the past, one is the most viable to compare with observations. In order to appreciate the origin of such EOS we identify the 
corresponding covariant energy-momentum tensor (EMT) of GDE from which the EOS arise. Such EMT's allow us to reveal some bad and good features inherent to the EOS.


\section{$f(R)$ theory, the Ricci scalar approach}
\label{sec:f(R)}
We consider a gravitational theory given by the following action
\begin{equation}
\label{f(R)}
S[g_{ab},{\mbox{\boldmath{$\psi$}}}] =
\!\! \int \!\! \frac{f(R)}{2\kappa} \sqrt{-g} \: d^4 x 
+ S_{\rm matt}[g_{ab}, {\mbox{\boldmath{$\psi$}}}] \; ,
\end{equation}
where  $\kappa \equiv 8\pi G_0$ (we use units where $c=1$), and $f(R)$ is a sufficiently smooth ({\rm i.e.} $C^3$) but otherwise arbitrary function of the 
Ricci scalar $R$. The first term corresponds to the modified gravity action, while the second is the usual action for the matter, where 
${\mbox{\boldmath{$\psi$}}}$ represents schematically the matter fields (including both the visible and possibly the dark matter).

The field equation arising from the action~(\ref{f(R)}) under the metric approach is
\begin{equation}
\label{fieldeq1}
f_R R_{ab} -\frac{1}{2}fg_{ab} - 
\left(\nabla_a \nabla_b - g_{ab}\Box\right)f_R= \kappa T_{ab}\,\,,
\end{equation}
where $f_R$ stands for $\partial_R f$, $\Box= g^{ab}\nabla_a\nabla_b$ is the covariant D'Alambertian and $T_{ab}$ is the energy-momentum tensor of matter 
resulting from the variation of the matter action in~(\ref{f(R)}). It is straightforward to write the above equation in the following way:
\begin{eqnarray}
\label{fieldeq2}
&& f_R G_{ab} - f_{RR} \nabla_a \nabla_b R - 
 f_{RRR} (\nabla_aR)(\nabla_b R) \nonumber \\
&+&  g_{ab}\left[\frac{1}{2}\left(Rf_R- f\right)
+ f_{RR} \Box R + f_{RRR} (\nabla R)^2\right]  = \kappa T_{ab}\,\,,\nonumber \\
\end{eqnarray}
where $G_{ab}= R_{ab}-g_{ab}R/2$ is the Einstein tensor and $(\nabla R)^2:= g^{ab}(\nabla_aR)(\nabla_b R)$. Taking the trace of this equation yields
\begin{equation}
\label{traceR}
\Box R= \frac{1}{3 f_{RR}}\left[\rule{0mm}{0.4cm}\kappa T - 3 f_{RRR} (\nabla R)^2 + 2f- Rf_R \right]\,\,\,,
\end{equation}
where $T:= T^a_{\,\,a}$. Finally, using~(\ref{traceR}) in~(\ref{fieldeq2}) we find
\begin{eqnarray}
\label{fieldeq3}
& G_{ab} =& \frac{1}{f_R}\Bigl{[} f_{RR} \nabla_a \nabla_b R +
 f_{RRR} (\nabla_aR)(\nabla_b R) \nonumber \\
&  & -\frac{g_{ab}}{6}\Big{(} Rf_R+ f + 2\kappa T \Big{)} 
+ \kappa T_{ab} \Bigl{]} \; .
\end{eqnarray}
Equations~(\ref{fieldeq3}) and~(\ref{traceR}) are the basic equations for $f(R)$ gravity that we introduced in~\cite{Jaime2011}  and that have 
been employed recently to treat other related problems~\cite{Jaime2012a, Jaime2012b, Jaime2013, Liddle2013}. 

An important property of $f(R)$ gravity is that the EMT of matter $T_{ab}$ is conserved. That is, one can prove that the field equations imply $\nabla^a T_{ab}=0$. 
This property incorporates the {\it weak equivalence principle} in the theory and will play an important role in the analysis of the EOS of GDE. 
Moreover, as it is well known, if $2f(R_1)- R_1f_R(R_1)=0$, for some $R=R_1= const.$, the theory gives rise to an effective cosmological constant 
$\Lambda_{\rm eff}:= R_1/4$. Therefore, when this $R_1$ exists, and when the matter content is small or negligible compared with 
$\Lambda_{\rm eff}/\kappa$, as it is the case with a late Friedman-Robertson-Walker cosmology, then the expansion naturally accelerates within the $f(R)$ gravity 
when $R\rightarrow R_1$.

In the present article we shall focus on three models:
\begin{itemize}

\item Starobinsky~\cite{Starobinsky2007}:
\begin{equation}
f(R)= R+\lambda R_{S}\left[ \left( 1+\frac{R^2}{R^2_{S}}\right)^{-q}-1\right] \; ,
\end{equation}
with $q=2$ and $\lambda = 1$ and $R_{S}=4.17H_{0}^{2}$.
\\
\item Hu--Sawicky~\cite{Hu2007}:
\begin{equation}
f(R)= R- R_{\rm HS}\frac{c_{1}\left(\frac{R}{R_{\rm HS}}\right)^n}{c_{2}\left(\frac{R}{R_{\rm HS}}\right)^n+1} \; ,
\end{equation}
where the parameters are $n=4$, $c_1\approx 1.25 \times 10^{-3} $, $c_2\approx 6.56 \times 10^{-5}$ 
and $R_{\rm HS}\approx 0.24 H_0^2$.
\\
\item Miranda {\it et al.}~\cite{Miranda2009} (hereafter MJW model):
\begin{equation}
f(R)= R-\beta R_{*}{\rm ln}\left( 1+\frac{R}{R_{*}}\right) \; ,
\end{equation}
with $\beta=2$ and  $R_{*}= H_0^2$.
\end{itemize}

All these models can be cosmologically viable (at least at the background level)~\cite{Jaime2012a}, and pass the Solar System tests. 
Nevertheless the MJW model has been the object of debate~\cite{delaCruz2009,Thongkool2009,Miranda2009b}.

\section{Cosmology}
\label{sec:cosmology}
We assume now a FRW metric,
\begin{equation}
\label{SSmetric}
ds^2 = - dt^2  + a^2(t)\!\left[ \frac{dr^2}{1-k r^2} + r^2 \left(d\theta^2 + \sin^2\theta d\varphi^2\right)\right]
\!,\!
\end{equation}
where $k=\pm 1,0$, and focus only on the ``flat'' case $k=0$. We also assume that the EMT of matter $T_{ab}$ is a mixture of three kinds of perfect 
fluids, $T_{ab}= \sum_{i=1}^3 T_{ab}^i$ (baryons, radiation and dark matter) in epochs where they do not interact with each other except gravitationally. 
So in our analysis we do not put forward $f(R)$ gravity as a possible solution to the dark matter problem but only as an alternative to the dark-energy. 
The total energy-density of matter is then $\rho= -T^{t}_{\,\,t}$ which is given by Eq.~(\ref{rhomatt}) below, 
with pressures $p_{\rm bar,DM}=0$ and $p_{\rm rad}= \rho_{\rm rad}/3$. 
Under these assumptions, Eqs.~(\ref{traceR}) and~(\ref{fieldeq3}) read,
\begin{eqnarray}
\label{traceRt}
& \ddot R = &-3H \dot R -  \frac{1}{3 f_{RR}}\left[ 3f_{RRR} \dot R^2 + 2f- f_R R \right. \nonumber\\
& &  - \kappa (\rho_{\rm bar}+ \rho_{\rm DM}) \Big]\,\,\,,\\
\label{Hgen}
& H^2 = &\frac{\kappa}{3}\left(\rule{0mm}{0.3cm} \rho +\rho_{X}\right) \,\,\,,\\
\label{Hdotgen}
& \dot{H}= & -H^2 -\frac{\kappa}{6}\left\{\rule{0mm}{0.4cm} \rho +\rho_{X}+3\left(p_{\rm rad}+ p_{X}\right) \right\} \,\,\,,\\
\label{Hubble}
& H = & \dot a/a \,\,\,,
\end{eqnarray}
where $\dot{}\,\,= d/dt$ and $\rho_{X}$, $p_{X}$ are the density and pressure of GDE, respectively, given explicitly by ${\tilde \rho}_X$ and 
${\tilde p}_X$ according to Table~\ref{tab:EOS}, taking $A=1=B$ (see Section~\ref{sec:EOS}).

\begin{table*}
\centering
\begin{tabular}{|c| c| c| c|}
\hline 
EMT of GDE & \multicolumn{3}{c|}{energy-density and pressure of GDE}  \\
\hline  
& \multicolumn{3}{l|}{ } \\
 & \multicolumn{3}{l|}{
${\tilde \rho}_X = \frac{A}{\kappa f_{R}}\left[\rule{0mm}{0.5cm} \frac{1}{2}\left( f_{R}R-f\right) -3f_{RR}H\dot{R} + 
\kappa \rho\left(1- \frac{B f_{R}}{A}\right) \right] $}     \\
${\tilde T}_{ab}^{\,X}(A,B):= A T_{ab}^{\rm tot} - B T_{ab}$ & \multicolumn{3}{c|}{} \\
& \multicolumn{3}{c|}{
${\tilde p}_X = -\frac{A}{3\kappa f_{R}}\left[\rule{0mm}{0.5cm}
 \frac{1}{2}\left(f_{R}R+f \right) + 3f_{RR}H\dot{R}-\kappa\left(\rho -3 p_{\rm rad} \frac{B f_R}{A}\right) \right]$} \\
& \multicolumn{3}{l|}{ } \\
\hline 
\multicolumn{1}{c}{} & \multicolumn{1}{c}{} & \multicolumn{1}{c}{} & \multicolumn{1}{c}{} \\  
\hline 
\pbox{6cm}{Definition} & $A$ & $B$ & EMT \\ 
\hline
& & & \\
 I ($T_{ab}^{X}$, $\rho_X$, $p_X$, $\omega_X$)  & $1$ & $1$  &  $T_{ab}^{X}$ conserved \\
\hline
& & & \\
  II ($T_{ab}^{II\,,\,X}$, $\rho_X^{II}$, $p_X^{II}$, $\omega_X^{II}$) & $f_R^0$ & $1$ & $T_{ab}^{II\,,\,X}$ conserved  \\
\hline
& & & \\
 III ($T_{ab}^{III\,,\,X}$, $\rho_X^{III}$, $p_X^{III}$, $\omega_X^{III}$) & $f_R$  & $1$  & $T_{ab}^{III\,,\,X}$ not conserved \\
\hline
& & & \\
 IV  ($T_{ab}^{IV\,,\,X}$, $\rho_X^{IV}$, $p_X^{IV}$, $\omega_X^{IV}$) & $1$    & $f_R^{-1}$  &  $T_{ab}^{IV\,,\,X}$ not conserved (conserved only in vacuum) \\
\hline 
\end{tabular}
\caption{Top: Geometric dark energy (GDE) variables in terms of the scalar functions $A$ and $B$. 
The EMT  $T_{ab}^{\rm tot}$ is given explicitly by the r.h.s of 
equation~(\ref{fieldeq3}) while $T_{ab}$ is the total EMT of matter. In the present case, $T_{ab}$ corresponds to a 
combination of three perfect fluids describing radiation, baryons and dark matter. 
The explicit expressions for ${\tilde \rho}_X$ and ${\tilde p}_X$ displayed in the table can be obtained from ${\tilde T}_{ab}^{\,X}(A,B)$.  
In the GR case [$f(R)= R$] all the GDE quantities vanish identically, as expected, even if the matter terms appear explicitly in 
these expressions (i.e. the GDE contributions ensue only when the theory departs from GR). 
Bottom: The different definitions of the EMT, energy-density, pressure and EOS of GDE are obtained from the quantities of top panel 
taking the parameters $A$ and $B$ as indicated in the table. Definitions III and IV produce the same EOS since $B f_R/A=1$ in both cases. 
The EMT $T_{ab}^{X}$ and $T_{ab}^{II\,,\,X}$ are conserved, while $T_{ab}^{III\,,\,X}$ and $T_{ab}^{IV\,,\,X}$ are not.}
\label{tab:EOS}
\end{table*}

\begin{figure}
\includegraphics[width=8.5cm,height=4cm]{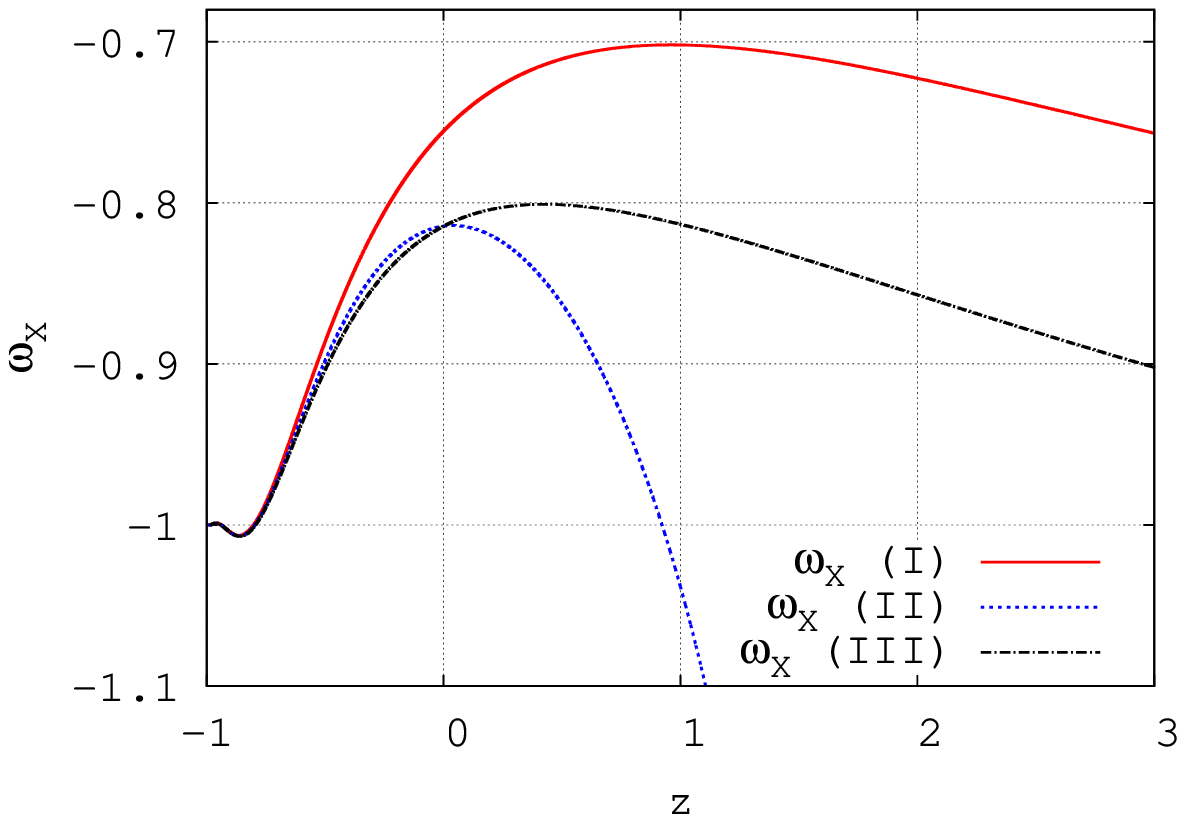}

\includegraphics[width=8.5cm,height=4cm]{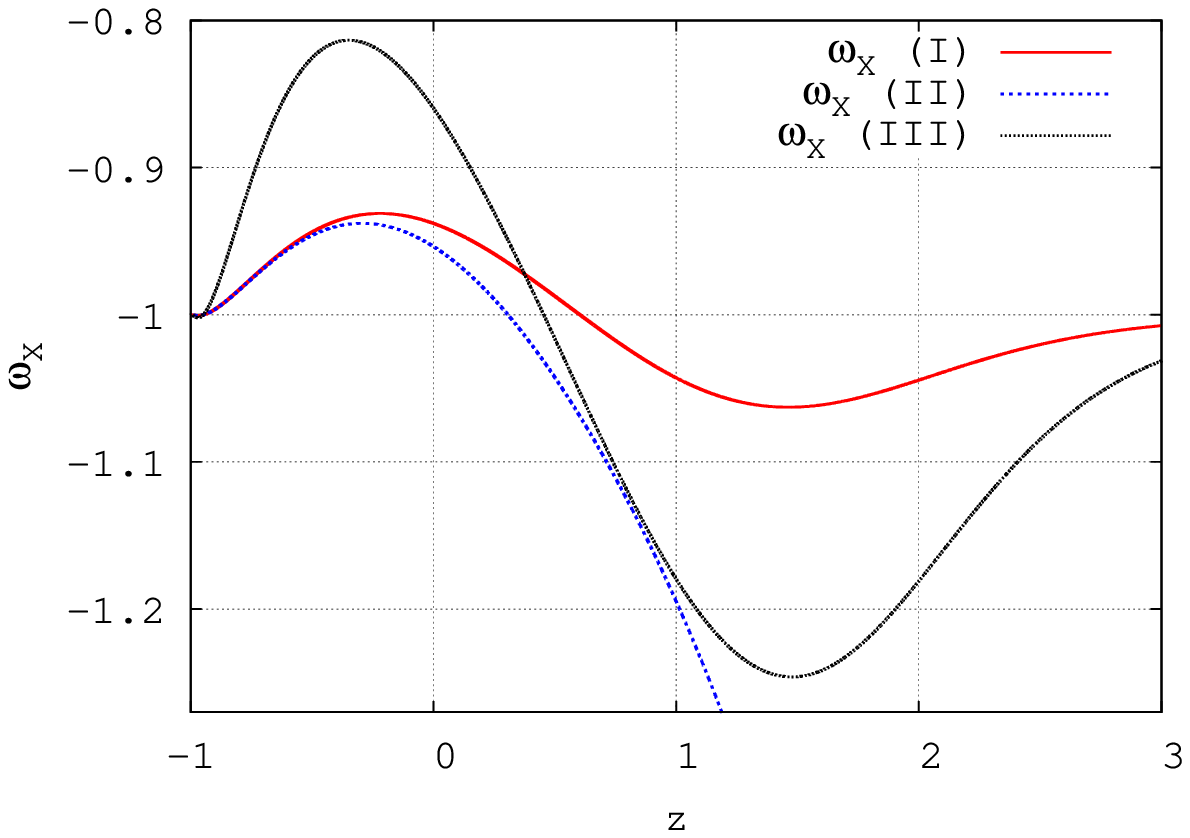} 

\includegraphics[width=8.5cm,height=4cm]{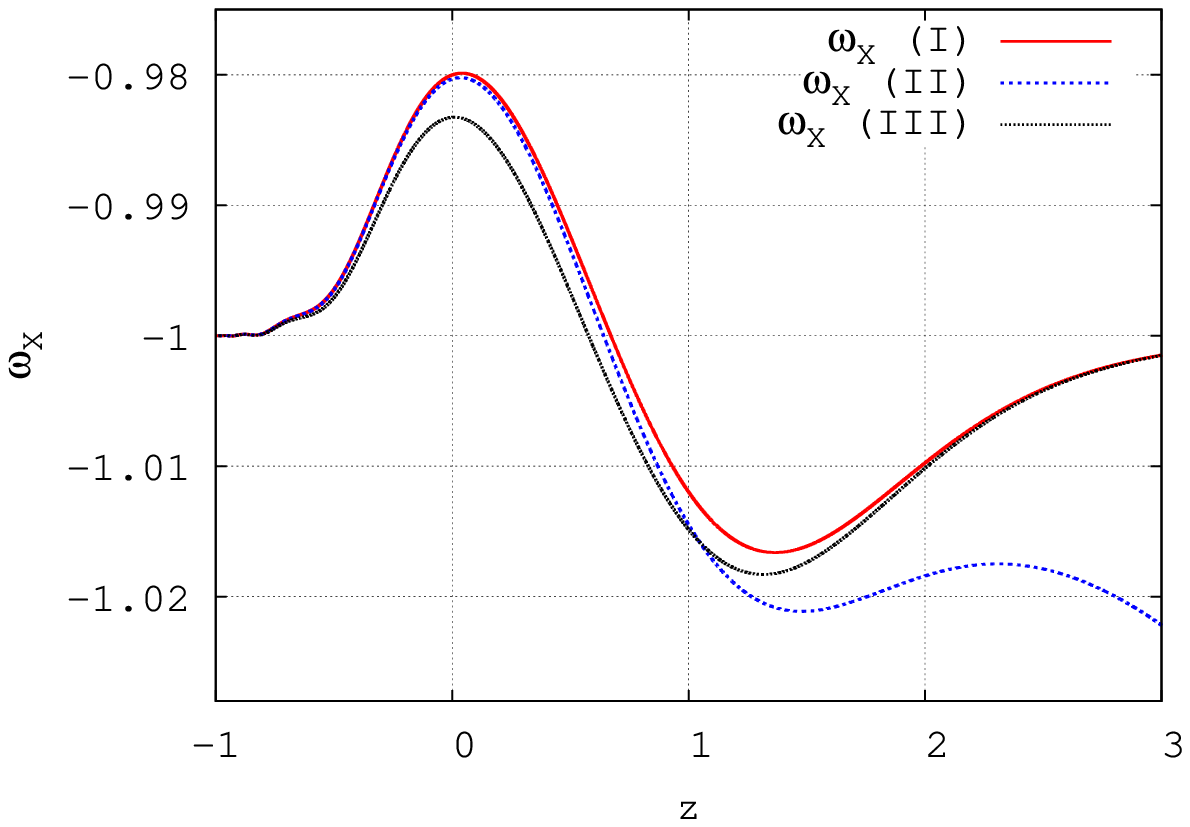}

\caption{(color online). Equations of state $\omega_X$, $\omega_X^{II}$ and $\omega_X^{III}$ as defined in the main text (denoted by $\omega_X (I)$, 
$\omega_X (II)$ and $\omega_X (III)$ in the panels) as a function of $z=a_0/a-1$ for the models MJW (left panel), Starobinsky (middle panel) and 
Hu--Sawicky (right). We appreciate the nonequivalence between the three EOS in each model. In the far future all the EOS coincide since the $f(R)$ 
model is dominated by an effective cosmological constant which is associated with the de Sitter point $R_1=4\Lambda_{\rm eff}$.}
\label{fig:wxall}
\end{figure}

As usual, the matter variables obey their own dynamics provided by $\nabla_a T^{ab}=0$ which leads to the well known conservation equation 
\begin{equation}
\label{consfluid}
\dot \rho_i +3 H \left(\rho_i + p_i\right) =0 \,\,\,,
\end{equation}
($i=1-3$) for each fluid component. This equation integrates straightforwardly for each of those species and the total energy-density of matter reads 
\begin{equation}
\label{rhomatt}
\rho= (\rho_{\rm bar}^0 + \rho_{\rm DM}^0)(a/a_0)^{-3} + \rho_{\rm rad}^0 (a/a_0)^{-4}\,\,\,,
\end{equation}
where the knotted quantities stand for the values today. 
The quantities $\rho_{X}$ and $p_{X}$ also satisfy a conservation equation similar to~(\ref{consfluid}), but with an EOS 
$\omega_{X}$ given by Eq.~(\ref{EOSX1c}) below, which evolves in cosmic time (cf. Figure~\ref{fig:wxall}). 
The {\it total} equation of state is defined by $\omega_{\rm tot} := p_{\rm tot}/\rho_{\rm tot}$, where $p_{\rm tot}$ is the total pressure, and 
$\rho_{\rm tot}$ is the total energy density, including the contributions of the geometric-dark components: 
\begin{equation}
\label{EOSTOT}
\omega_{\rm tot} = -\frac{1}{3}\left[ \frac{\frac{1}{2}\left(f_{R}R+f \right) + 3f_{RR}H\dot{R}-\kappa\rho}
          {\frac{1}{2}\left( f_{R}R-f\right) -3f_{RR}H\dot{R} + \kappa \rho}\right] \,\,\,.
\end{equation} 

This quantity is directly related with the deceleration parameter 
$q := -\frac{\ddot{a}}{aH^2}= \frac{1}{2}\left(1+3\omega_{\rm tot}\right)$.

Now, the expression for the Ricci scalar computed directly from the metric is given by
\begin{equation}
\label{Rt}
R= 6\left(\dot H + 2 H^2 + \frac{k}{a^2}\right) \,\,\,.
\end{equation}

The EOS of geometric dark energy is given by $\omega_{X} := p_X/\rho_X$ (cf. Table~\ref{tab:EOS}) and can be conveniently written 
in the following way using Eqs.~(\ref{Rt}) (with $k=0$) and (\ref{Hdotgen})
\begin{equation}
\label{EOSX1c}
\omega_{X} := \frac{1-\Omega_{\rm rad} -\frac{R}{3H^2}}{3 \Omega_X}\,\,\,,
\end{equation}
where the dimensionless densities for the different species $\Omega_i:= \kappa \rho_i/(3H^2)$ (the subindex $i$ stands now for radiation, 
baryons, DM and GDE)~\footnote{Some authors~\cite{Amendola2007b,Tsujikawa2007,Amendola2008,Tsujikawa2008,Gannouji2009} define $\Omega_i= \kappa \rho_i/(3 f_R H^2)$. 
This difference must be borne in mind when comparing results.} satisfy the constraint $\Omega_{\rm rad} + \Omega_{\rm bar}+ \Omega_{\rm DM} + \Omega_X= 1$. 
Clearly we assume here that $f(R)\neq R$ as otherwise one is led to $\omega_{X}=0/0$.

We have solved the system~(\ref{traceRt}),~(\ref{Hdotgen}) and~(\ref{Hubble}) as well as the alternative 
system~(\ref{traceRt}),~(\ref{Hubble}) and~(\ref{Rt}) using a fourth order Runge--Kutta 
algorithm integrating from the matter or radiation dominated epochs to the present (see ~\cite{Jaime2012a} for the details). We checked that 
both systems give the same results.


\section{Alternative EOS of GDE}
\label{sec:EOS}
One of the goals of physical cosmology is to measure with better accuracy the EOS for dark energy $\omega_{\rm DE}$~\cite{BB,obsEOS}. A departure 
from unit of $|\omega_{\rm DE}|$ would indicate that its nature is in fact much more complex than a ``simple'' cosmological constant. In the framework 
of $f(R)$ gravity, unfortunately there is some kind of ambiguity in the way the EOS of GDE can be defined. For instance, the definition given by some 
authors depend on the way they decided to split both the total (effective) energy-density that appears in~(\ref{Hgen}) and the total pressure that 
appears in~(\ref{Hdotgen}). This ambiguity amounts to disentangle this EOS from $\omega_{\rm tot}$, which in $f(R)$ gravity, seems {\it a priori} 
a non trivial task. Nevertheless, it is indeed possible to analyze if some definitions are better motivated than others, or if some of them have 
inherent pathologies. The definitions displayed in the previous section are those that we consider having the best features and are marked 
below as Definitions I. 

It is convenient to introduce the covariant EMT's from which the different definitions can be deduced. Let us start then with 
$\rho_{X}$ and $p_{X}$ given previously. As one can verify, these emerge from an EMT defined from~(\ref{fieldeq3}) as follows 
(cf. Refs.~\cite{Starobinsky2007,Hu2007b}):
\begin{equation}
\label{EMTIa}
T_{ab}^{X}:= T_{ab}^{\rm tot} - T_{ab} \,\,\,,
\end{equation}
where $T_{ab}^{\rm tot}$ is read-off from the r.h.s of~(\ref{fieldeq3}) which states
\begin{equation}
\label{Einst}
G_{ab}= \kappa T_{ab}^{\rm tot} \,\,\,.
\end{equation}
The EMT $T_{ab}^{X}$ satisfies the conservation equation $\nabla^a T_{ab}^{X}=0$ by virtue of the Bianchi identities and because $T_{ab}$ is conserved as well, 
as we stressed in Section~\ref{sec:f(R)}. 

At first sight it may seem awkward to see the EMT of matter $T_{ab}$ appearing in the definition of $T_{ab}^{X}$~(\ref{EMTIa}), which is supposed to be 
described only in terms of geometrical quantities. Nonetheless, there is a simple way to rewrite $T_{ab}^{X}$ in that way as follows:
\begin{equation}
\label{EMTIb}
T_{ab}^{X}=  \kappa^{-1}\left(G_{ab}- \mathfrak{G}_{ab}\right) \,\,\,,
\end{equation}
where $\mathfrak{G}_{ab}$ is given by the l.h.s of (\ref{fieldeq2}). This is certainly a more complicated alternative but otherwise equivalent to Eq.~(\ref{EMTIa}). 
For instance, notice that in this alternative expression (\ref{EMTIb}) we will find terms like $\Box R$ that in (\ref{EMTIa}) are rewritten using (\ref{traceR}). 
We opted for (\ref{EMTIa}) over (\ref{EMTIb}) since the former suits better our approach, but again, both are equivalent. So the reader should not get the wrong idea 
that the appearance of the matter terms in (\ref{EMTIa}), and therefore in $\rho_{X}$ and $p_{X}$ as given by Table~\ref{tab:EOS}, 
imply a direct coupling (i.e. other than gravitational) 
between the matter and the geometric dark energy as if both, $T_{ab}$ and $T_{ab}^{X}$ were not conserved individually. They are, by construction. 

Equivalent expressions for $T_{ab}^{X}$ based on (\ref{EMTIb}) can be found in Refs.~\cite{Starobinsky2007,Motohashi2010,Motohashi2011a,
Motohashi2011b,Lee2011}. In particular, in~\cite{Miranda2009,deFelice2010,Motohashi2010,Motohashi2011a,Motohashi2011b,Bamba}, one can find a definition 
for the EOS which is equivalent to our $\omega_X$.

Alternatively, one can consider other definitions for the geometric dark components. For instance, ${\tilde \rho}_X$ and ${\tilde p}_X$ 
given in Table~\ref{tab:EOS} encompass all the expressions found in the literature according 
to the different values assigned to the scalar functions $A$ and $B$, as we 
will show next. These alternate expressions for the density and pressure of GDE are derived from an EMT given by 
\begin{equation}
\label{EMTX}
{\tilde T}_{ab}^{\,X}(A,B):= A T_{ab}^{\rm tot} - B T_{ab} \,\,\,, 
\end{equation}
which can be written in terms of purely geometrical quantities as 
\begin{equation}
\label{EMTXb}
{\tilde T}_{ab}^{\,X}(A,B)= \kappa^{-1}\left(AG_{ab}- B\mathfrak{G}_{ab}\right) \,\,\,. 
\end{equation}

\noindent{\bf Definitions I.} We labeled them simply by $\rho^X$ and $p^X$ and are given explicitly by 
${\tilde \rho}_X$ and ${\tilde p}_X$ taking $A=B=1$ from Table~\ref{tab:EOS}. Both $\rho^X$ and $p^X$ are associated with $T^X_{ab}$ (\ref{EMTIa}). 
Alternatively if one uses (\ref{EMTXb}) with the previous values for $A$ and $B$ one obtain the alternative expressions for $\rho^X$ and $p^X$ without 
the matter terms (see~\cite{Miranda2009}).
\smallskip 

\begin{figure}
\includegraphics[width=8.5cm,height=4cm]{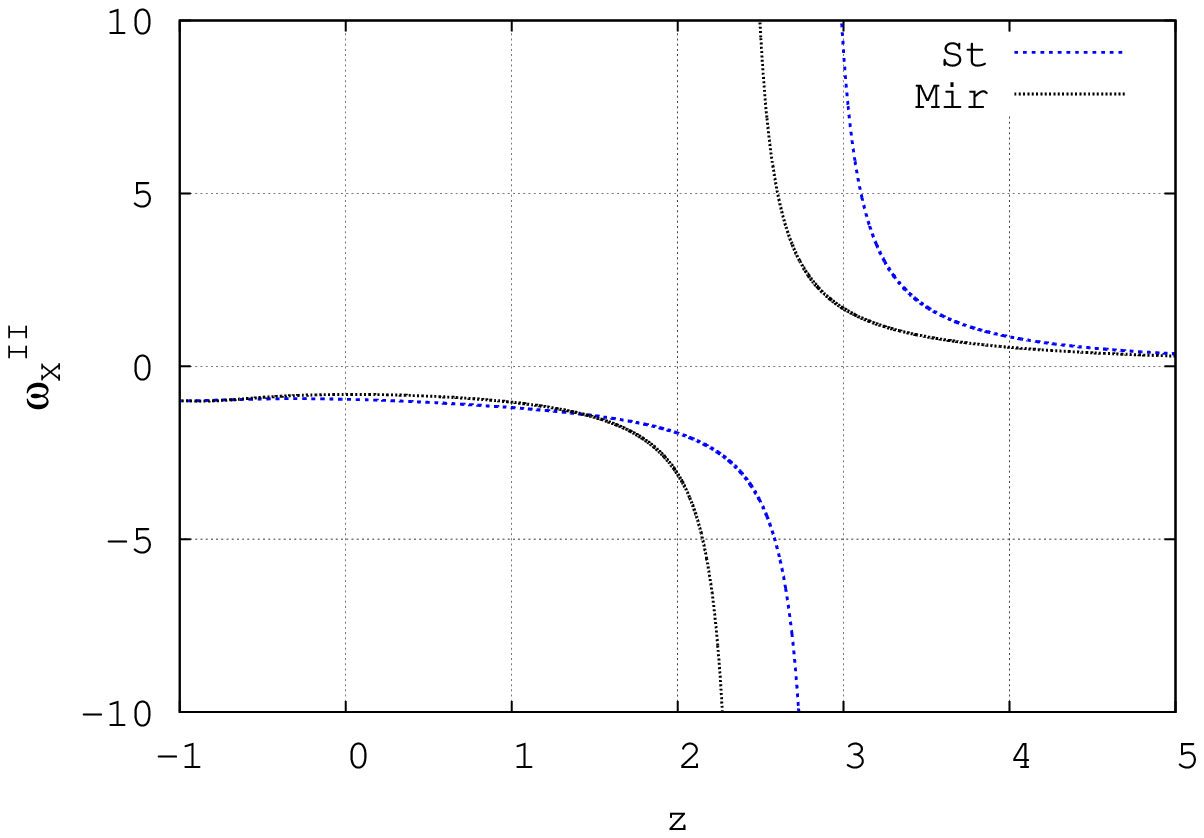} 

\includegraphics[width=8.5cm,height=4cm]{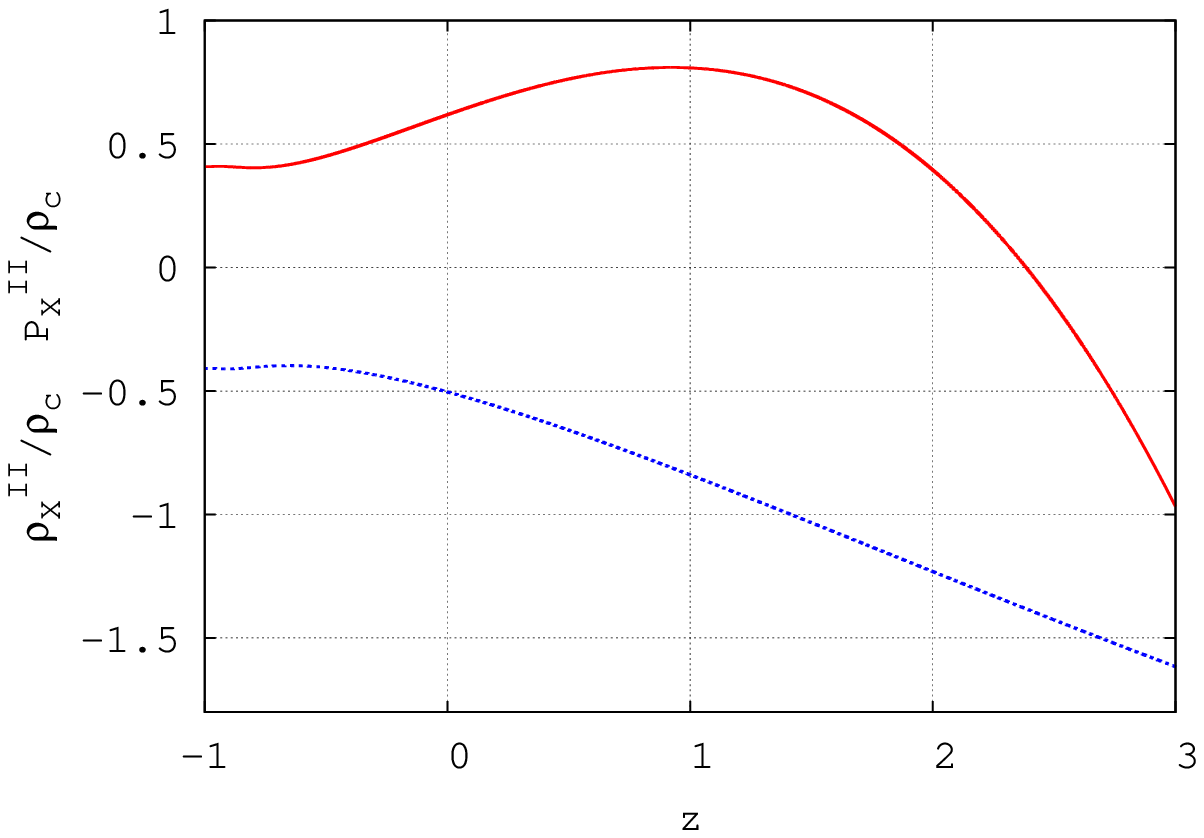}  

\includegraphics[width=8.5cm,height=4cm]{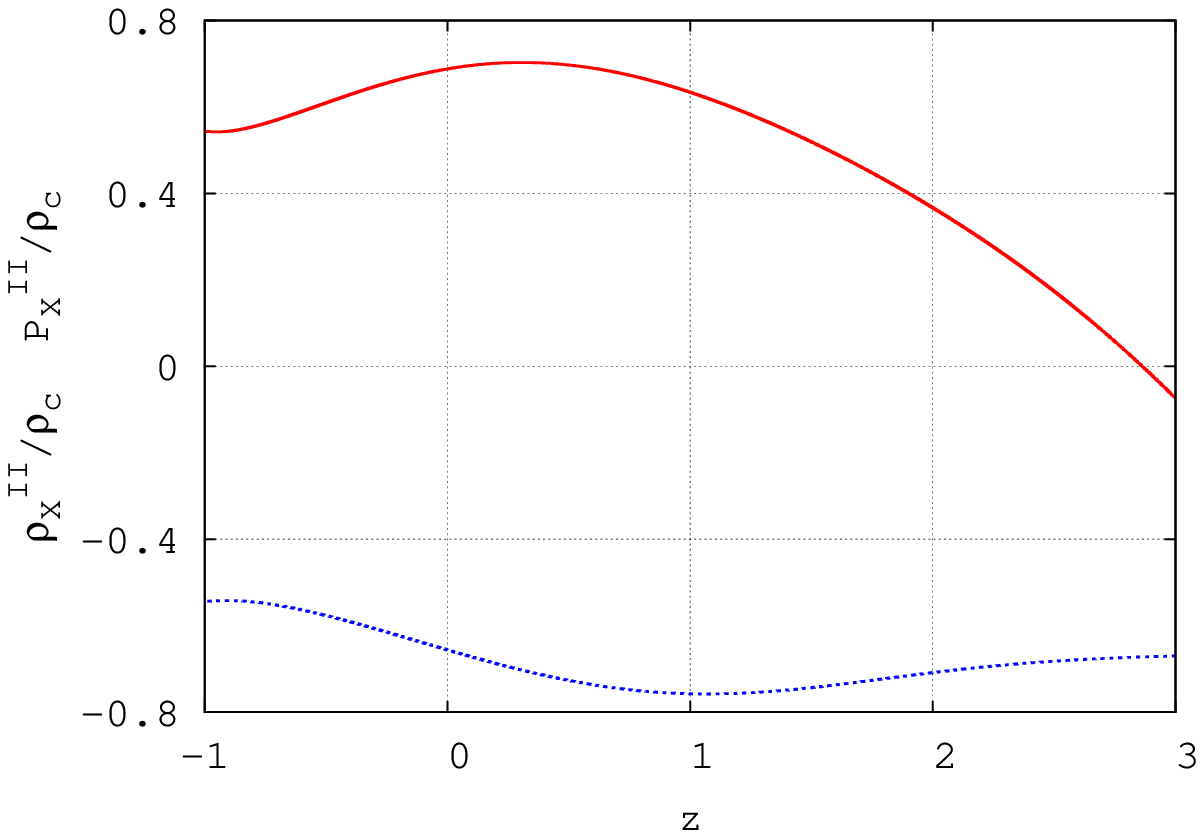}
\caption{(color online). Left: Equation of state $\omega_X^{II}$ for the Starobinsky and MJW models. Notice the divergence due to the vanishing of 
$\rho_X^{II}$ (see the other panels). For these two $f(R)$ models the EOS is also plotted in figure~\ref{fig:wxall}. Middle: Energy-density 
$\rho_X^{II}$ (solid line) and pressure $p_X^{II}$ (dotted line) computed from the MJW model. The quantities are given in units of 
$\rho_{\rm c}^0= 3 H_0^2/(8\pi G_0)$. Notice that $\rho_X^{II}$ becomes zero at $z\approx 2.38$. Right: same as middle panel but using the Starobinsky 
model. In this case $\rho_X^{II}$ becomes zero at $z\approx 2.86$.}
\label{fig:wx2-St-Mir}
\end{figure}

\noindent{\bf Definitions II.} These are given by taking $B=1$ and the {\it ad-hoc} choice $A=f_R^0= const.$ (i.e. $f_R$ today; see~\cite{Gannouji2009} 
for a discussion concerning this constant $f_R^0$) leading to the GDE variables introduced in~\cite{Amendola2008,Amendola2007b,Tsujikawa2007,Gannouji2009}.
Like $T_{ab}^{X}$, the EMT $T_{ab}^{II\,,\,X}$ is also conserved since $f_R^0$ is just a constant. The energy-density ${\rho^{II}_X}$, pressure $p^{II}_X$ and 
EOS $\omega^{II}_X := p^{II}_X/\rho^{II}_X$ for this second definition are obtained taking $A=f_R^0$ and $B=1$ in Table~\ref{tab:EOS}. 
In general $f_R^0 \neq 1$ although $f_R^0 \approx 1$ (cf. Ref.~\cite{Hu2007}) 
\footnote{Since this dimensionless quantity appears as $G_{\rm eff}= G_0/f_R$ and at present time $G_{\rm eff}^0\approx G_0$, 
thus one expects $f_R^0 \approx 1$.}. As shown by Hu \& Sawicky~\cite{Hu2007}, one has to take (at least in their model) $f_R^0 \approx 1$ 
in order to avoid large deviations in the EOS relative to $\omega_\Lambda=-1$. However, if one takes $f_R^0 \neq 1$, $\omega_X^{II}$ and 
$\omega_X$ are not equivalent, and furthermore $\omega_X^{II}$ can diverge at some redshift depending on the $f(R)$ model (see Figure~\ref{fig:wx2-St-Mir}). 
The divergence of this EOS is a drawback that was already observed 
in~\cite{Amendola2008} and later discussed by Starobinsky~\cite{Starobinsky2007}. Due to this pathological behaviour EOS $\omega_X^{II}$ 
is not suitable for comparing with observations.
\smallskip

\noindent{\bf Definitions III.} The third kind of definitions are associated with the choice $A=f_R$ and $B=1$. 
This is similar to $T_{ab}^{II\,,\,X}$ except that now $f_R$ is not a constant. 
This EMT was considered by Sotiriou \& Faraoni~\cite{Sotiriou2010} (denoted $\,T_{\mu\nu}^{(eff)}$ by them). 
The quantities $\rho^{III}_X$ and $p^{III}_X$ are obtained from $T_{ab}^{III\,,\,X}$ (see Table~\ref{tab:EOS}). It is interesting to note that for the vacuum solution 
$R=R_1$ discussed previously, these definitions read:
\begin{equation}
\rho^{III}_X = f_R \Lambda_{\rm eff} \kappa^{-1} \,\,\,,\,\,\,p^{III}_X = - f_R \Lambda_{\rm eff} \kappa^{-1} \,\,\,,
\end{equation}
which, due to the presence of the factor $f_R$, are not related to $\Lambda_{\rm eff}$ like $\rho_X=\Lambda_{\rm eff}/\kappa$ and 
$p_X=-\Lambda_{\rm eff}/\kappa$. This already shows the unnatural character of Definitions III, despite 
the fact that the factor $f_R$ cancels out when considering the EOS $\omega^{III}_X:= p^{III}_X/\rho^{III}_X=-1$. 
Notwithstanding, the most disturbing aspect of this third definition is that in the realistic scenario when matter is present and where $R$ varies with the cosmic 
time, the EMT $T_{ab}^{III\,,\,X}$ is not conserved because the scalar $f_R$ is not a constant, and so
\begin{equation}
\nabla^a T_{ab}^{III\,,\,X} = T_{ab}^{\rm tot} \nabla^a f_R \neq 0 \,\,\,.
\end{equation}
\smallskip 

\noindent{\bf Definitions IV.} The fourth type of EMT for GDE is obtained from the choice $A=1$ and $B= f_R^{-1}$ in (\ref{EMTX}). 
This EMT leads to the quantities $\rho^{IV}_X$ and $p^{IV}_X$ (see Table~\ref{tab:EOS}). We find the following relationships between Definitions III and IV,
\begin{equation}
{\rho^{IV}_X} =  \frac{\rho^{III}_X}{f_R} \,\,\,, \,\,\,p^{IV}_X =   \frac{p^{III}_X}{f_R} \,\,\,,\,\,\,
\omega^{IV}_X \equiv  \omega^{III}_X \,\,\,.
\end{equation}
Definitions IV has been considered by several authors~\cite{Capozziello2005,Capozziello2008a, Capozziello2008b,Carloni2009,Clifton2012,Bamba,Bamba2012,Lee2011}. 
Like $T_{ab}^{III\,,\,X}$, $T_{ab}^{IV\,,\,X}$ is not conserved either. 
Only in vacuum $T_{ab}^{IV\,,\,X}$ is conserved as it coincides with $T_{ab}^{X}$, and in that instance the EOS $\omega^{IV}_X$ and $\omega^{III}_X$ 
reduce to $\omega_X$. Nonetheless, since the Universe does contain matter, Definitions I are different from III and IV in general, and so $\omega_X$ is not equivalent 
neither to $\omega_X^{III}$ nor to $\omega_X^{IV}$. Since both $T_{ab}^{III\,,\,X}$ and $T_{ab}^{IV\,,\,X}$ are not conserved, the corresponding densities and pressures 
will not obey an equation like (\ref{consfluid}). In particular
\footnote{cf. Eq.~(108) of Ref.~\cite{Capozziello2008a}, Eq.~(8) of Ref.~\cite{Capozziello2008b}, 
Eq.~(13) of Ref.~\cite{Capozziello2005} and Eq.~(11.3) of Ref.~\cite{Bamba2012}.} 
\begin{equation}
\label{dotrho4}
\dot \rho^{IV}_X + 3 H \left(\rho^{IV}_X + p^{IV}_X\right) = \rho \,\frac{d (f_R^{-1})}{dt} \neq 0\,\,\,.
\end{equation}
The additional source term that appears in the r.h.s of Eq.~(\ref{dotrho4})
depends on $\rho$ [see Eq.~(\ref{rhomatt})], and on the time variation of the 
scalar $f_R^{-1}$. Both never vanish in general.

\section{Discussion}

Figure~\ref{fig:wxall} shows the three EOS $\omega_{X}$, $\omega_{X}^{II}$, $\omega^{III}_X$. 
As concerns $\omega_{X}$, our results are consistent with those reported in 
Refs.~\cite{Hu2007,Miranda2009,Motohashi2010,Motohashi2011a,Motohashi2011b,Bamba}. 
The EOS $\omega_{X}$ and $\omega^{III}_X$, although inequivalent, give similar results for the three models. 
However, as we pointed out before, the EOS $\omega_{X}^{II}$ can have a completely different behavior as it diverges at some redshift 
(see Figure~\ref{fig:wx2-St-Mir}). This is because $\rho_X^{II}$ vanishes at that redshift and then becomes negative, as seen in 
Figure~\ref{fig:wx2-St-Mir} (middle and left panels). The vanishing of $\rho_X^{II}$ can be understood by looking at ${\tilde \rho}_X$ in 
Table~\ref{tab:EOS}, notably at the term $\rho (1-f_RB/A)$ with $B=1$, as we argue next. First, when $A=1$, which corresponds to $\rho_X$, 
the combination $\rho (1-f_R)$ is always positive since $0<f_R<1$ during the cosmic evolution. 
If the other contributions are positive, which is expected since $f_{RR}\ll 1$ and since the other terms will give rise to $\Lambda_{\rm eff}>0$, 
then $\rho_X>0$. Nevertheless when we consider $\rho_X^{II}$ ($A=f_R^0$), and if $f_R^0<1$, then the term $\rho (1-f_R/f_R^0)$ can be negative in epochs where 
$f_R>f_R^0$, which are precisely the epochs of large $R$ and large $z$ where $f_R\rightarrow 1$. In the matter domination era the negative 
term $\rho (1-f_R/f_R^0)$ can dominate over the geometric terms of $\rho_X^{II}$. This is exactly what happens as depicted in Figure~\ref{fig:wx2-St-Mir} 
(middle and left panels). As the evolution proceeds, $\rho$ decreases to a point where the other terms of $\rho_X^{II}$, which increase, balance exactly 
the negative contribution to give $\rho_X^{II}=0$. The evolution goes on to a point where $f_R<f_R^0$ or where $\rho$ is sufficiently small, 
and then $\rho (1-f_R/f_R^0)$ becomes positive or small, in which case $\rho_X^{II}$ becomes positive. 
All this behavior exacerbates when $f_R^0$ departs sufficiently from unity, since in that case the term $\rho (1-f_R/f_R^0)$ becomes even more relevant. 
This is what happens in the Starobinsky and MJW cases as the parameter $f_R^0$ is not fixed beforehand but it is rather predicted from the 
initial conditions in the past. Conversely, the Hu--Sawicky model was constructed in such a way that $f_R^0\approx 1$ and so $\rho (1-f_R/f_R^0)$ is less 
important at the same epochs given that $1-f_R/f_R^0\approx 0$. This explains why we did not encounter that divergence in this model in the range of the 
redshifts we explored. However, farther in the past where $\rho$ dominates over the geometrical terms, $\rho (1-f_R/f_R^0)$ might become important and 
can make $\rho_X^{II}$ to vanish in that model too. This depends on how much $1-f_R/f_R^0$ and $\rho$ increase and decrease respectively as looking from 
past to present. At any rate, this divergence is unacceptable and indicates that EOS $\omega_{X}^{II}$ is ill-defined.

In summary, there exist at least three inequivalent definitions $\omega_{X}$, $\omega_{X}^{II}$, $\omega^{III}_X$ that we have identified in the literature 
as representing the geometric dark energy component. Definitions III and IV are flawed as their respective EMT's are not conserved. In particular, the non 
conservation of $T_{ab}^{IV\,,\,X}$ manifests a spurious ``interaction'' between matter and GDE, which can be made to vanish if Definitions I or II are 
adopted. Notwithstanding, Definitions II seems rather {\it ad hoc}, and furthermore $\omega_{X}^{II}$ is prone to diverge during the cosmic evolution. 
We conclude that the most satisfactory definition for the dark energy in the framework of $f(R)$ gravity is provided by Definitions I. It is important to 
come to an agreement on which of those definitions will be ultimately compared with the cosmological observations in view of the forthcoming projects 
planed to determine with better precision the EOS of dark energy~\cite{BB,obsEOS}. 


\acknowledgments
This work was partially supported by grants DGAPA--UNAM IN107113 and SEP--CONACYT 166656.

\vskip -.5cm

\end{document}